# Elektrikli Kara Araçları İçin Doğrulama Protokollü Güvenli Şarj ve Ödeme Sistemi

ÖMER AYDIN

Dokuz Eylül Üniversitesi, İzmir, Türkiye
omer.aydin@deu.edu.tr

## *Özet*

Fosil yakıtların sınırlı kaynak olması, ekonomik ve çevresel olumsuzlukları göz önünde bulundurulduğunda gelecekte yerini başka enerji kaynaklarına bırakacağı aşikârdır. Fosil yakıtların yerini almaya aday kaynakların içinde elektrik ön plana çıkmaktadır. Yakın gelecekte elektrikli kara, hava ve deniz araçları gündelik hayatta daha çok yer almaya başlayacaktır. Bu nedenle bu cihazların şarj sistemleri ve şarj sonrası ödeme işlemleri için sistemler geliştirilmeye başlanmıştır. Bu konuda genel bir standart henüz oluşmamıştır. Bu çalışmada elektrikli kara araçlarında kullanılmak üzere bilinen siber saldırılara karşı güvenli, mahremiyeti ön planda tutan şarj ve ödeme sistemi önerilmiştir. Şarj cihazı ile aracın karşılıklı, kablolu veya kablosuz olarak birbirini bir doğrulama protokolü ile doğruladığı, veri iletişiminin şifreli olarak yapıldığı, ödeme işlemlerinin ise güvenli olarak gerçekleştirilerek araç sahiplerine faturalandırılan bir sistem önerilmiştir.

**Anahtar Kelimeler:** Elektrikli Kara Aracı, Doğrulama Protokolü, Şifreleme, Güvenlik, Ödeme Sistemi, Şarj Sistemi

# Secure Charging and Payment System for Electric Land Vehicles with Authentication Protocol

### *Abstract*

It is obvious that fossil fuels are a limited resource and will be replaced by other energy sources in the future considering economic and environmental problems. Electricity comes to the forefront among the sources that are candidates to replace fossil fuels. In the near future, electric land, air and sea vehicles will start to take more place in daily life. For this reason, systems for the charging systems of these devices and post-charge payments have been developed. There is no general standard on this issue yet. In this study, a charge and payment system, which is safe against known cyber-attacks for use in electric land vehicles, and which prioritizes privacy, is proposed. A system has been proposed to verify each other wired or wirelessly with an authentication protocol, where the data communication is encrypted, and the payment transactions are performed securely and invoiced to the vehicle owners.

## GİRİŞ

Elektriğin insanlar tarafından kullanılmaya başlamasından günümüze bu enerji birçok teknolojik gelişmenin ana kaynağı olmuştur. Televizyon, telefon, bilgisayar vb. birçok teknolojik buluş ve cihazın çalışması için gerekli enerji elektriktir. Bu gelişim sürecinde elektriğin diğer enerji kaynakları içinde kullanımı her geçen gün artmıştır. Nicholas Cugnot tarafından 1769 yılında buhar ile çalışan ilk kara aracının (Reitze Jr, 1977) geliştirilmesinden bugüne kara, deniz, hava ve uzay araçlarında ciddi değişimler meydana gelmiştir. İçten yanmalı motorlarda kullanı-lan fosil yakıtların çevre kirliliği vb. kötü etkileri, dünya üzerindeki miktarlarının kısıtlı olması ve yakın zamanda tükeneceği öngörüsü ile farklı enerji kaynaklarına yönelme ihtiyacı doğmuştur.

**Şekil 1.** *Türlerine Göre Fosil Yakıt Rezervlerinin Kalan Ömürleri (ETKB, 2017)*

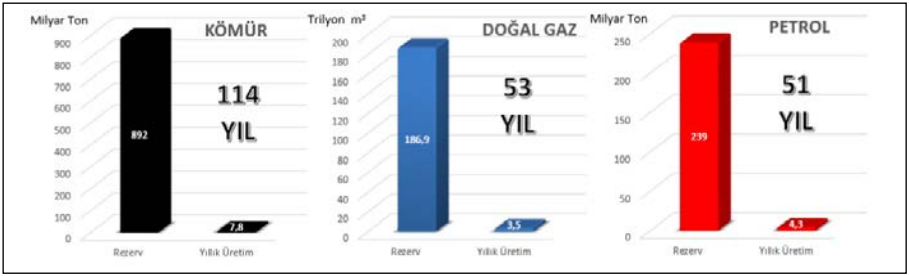

Fosil yakıt rezervleri hızla azalmakta olup özellikle petrol ve doğal gaz rezervleri kritik seviyelere yaklaşmaktadır (Öztornacı, 2019). Şekil 1'de dünya kömür, doğal gaz ve petrol rezervlerine ilişkin bilgi verilmiştir. Kaynaklara dağılım bakımından yıllara göre enerji talebinin dağılımı Tablo 1'de verilmiştir.

*Tablo 1. Birincil Enerji Talebinin Kaynaklara Dağılımı (Milyon Tona Eşdeğer Petrol) (Kalkınma Bakanlığı Özel İhtisas Komisyonu Raporu, 2014)*

|  | 1990 | 2010 | 2015 | 2020 | 2030 | 2035 |
|---|---|---|---|---|---|---|
| **Kömür** | 2.231 | 3.474 | 3.945 | 4.082 | 4.180 | 4.218 |
| **Petrol** | 3.230 | 4.113 | 4.352 | 4.457 | 4.578 | 4.656 |
| **D.Gaz** | 1.668 | 2.740 | 2.993 | 3.266 | 3.820 | 4.106 |
| **Nükleer** | 526 | 719 | 751 | 898 | 1.073 | 1.138 |
| **Hidrolik** | 184 | 295 | 340 | 388 | 458 | 488 |
| **Biyokütle** | 903 | 1.277 | 1.408 | 1.532 | 1.755 | 1.881 |
| **Diğer** | 36 | 112 | 200 | 299 | 554 | 710 |
| **Toplam** | 8.779 | 12.730 | 13.989 | 14.922 | 16.417 | 17.197 |

Tüm bu bilgileri göz önünde bulundurduğumuzda görüyoruz ki elektrik enerjisi ile birlikte hidrojen, biyolojik yakıtlar, nükleer reaksiyonlar enerji kaynakları kullanılmaktadır. Kullanım kolaylığı, erişilebilirliği ve üretim anlamında çeşitli avantajları da göz önünde bulundurulduğunda elektrik enerjisi günlük yaşamamızda kullanılan araçların kullandığı enerji kaynakları arasında ön plana çıkmaktadır. Elektrikli araçların kullanımı 19. yüzyıl sonları ve 20. yüzyıl ilk dönemlerinde başlamasına rağmen elektrikli araçlar, 1980 'den sonra ve özellikle 2000 'li yılların başından itibaren gelişerek ve yaygınlığı artarak günümüz dünyasında daha fazla yer almaya başladı. Elektrikli bisikletler, arabalar, tren, otobüs ve uçaklar artık ciddi anlamda yaygınlaşmaya başlamıştır. Bu araçlar içinde yer alan bireysel araç sınıfındaki elektrikli bisiklet ve otomobillerin bataryalarının doldurulması gerekliliği ve bu gerekliliğin getirdiği zorluklar çözülmesi gereken konular olarak karşımıza çıkmaktadır. Özellikle bireysel otopark imkânının çok düşük olduğu ülkelerde araçların şarjı için güvenli ve standart bir şarj sistemi ihtiyacı bulunmaktadır. Sokakta park etmiş araçların şarj edilmesi ve bu işlem yapılırken bilişim teknolojilerinden faydalanarak ücretlendirme, cihaz doğrulama, faturalandırma vb. konularda çalışacak bilgi sistemlerine ihtiyaç vardır. Bu yeni durum için oluşturulmuş bir standart bulunmamaktadır. Bu nedenle standardın oluşturulmasına yön verebilecek çalışmalar yapılması gereklidir.

Bu çalışmanın amacı elektrikli araç bataryalarının sokaklarda park halinde iken doldurulması için bir çözüm sunmaktır. Sunulan çözüm ile dolum işleminin ödemesinin alınması, cihazların karşılıklı doğrulama işlemini yapması, güvenli iletişimin sağlanması, dolum işleminin sonlandırılması ve faturalandırma için işlemler tanımlanacaktır. Bu tanımlar ve çözümler ile elektrikli araçların batarya dolum işlemleri için bir standart oluşturulmasına katkı sağlamak amaçlanmıştır.

## GÜVENLİ ÖDEME VE ŞARJ SİSTEMİ

Bireysel otopark alanlarının yeterli olmadığı ülkelerde sokaklarda, alışveriş merkezi otoparklarında ve diğer otopark alanlarında cihazların park halinde iken kişilerin araçlarını şarj edebilmeleri birçok avantaj sağlayacaktır. Özellikle birçok kişinin araçlarını akşam iş çıkışı eve dö-nüp sabah tekrar işe gidecekleri zamana kadar veya hafta sonları sokaklara park ettiği düşünüldüğünde elektrikli araçlarını bu sürede güvenli şekilde şarj edebilmeleri büyük önem kazanmaktadır. Bireysel otoparkı olmayan apartmanlarda araçlar, şarj süresince dışarıdan gelebilecek müdahalelere açıktır. Bu süreçte şarj bağlantısının kesilmesi, şarjın müdahale ile kesilmeden başka bir cihaza bağlanarak faturalandırmanın yanlış kişiye yapılması vb. birçok risk bulunmaktadır. Tüm bunlar nedeni ile aşağıda detaylarını paylaşılan doğrulama protokolü ve faturalandırma sistemi tasarlanmıştır.

**Şekil 2.** *Şarj Ve Ödeme Sistemi Bileşenlerinin Sokak Üzerinde Yerleşiminin Görünümü*

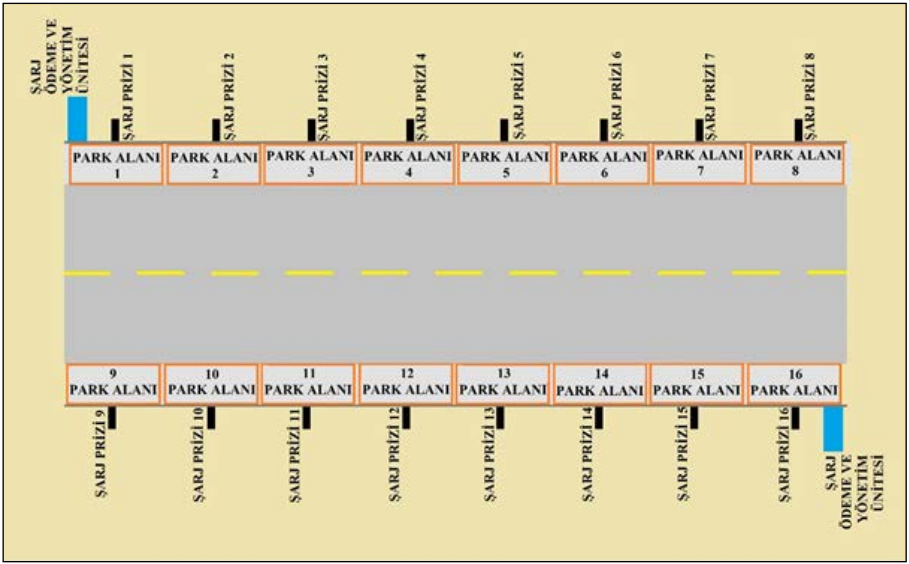

Sokaklara araçların park edilişi ve önerilen sokak üstü şarj ve faturalandırma sisteminin temsili görüntüsü Şekil 2'de gösterilmiştir. Araçların şekildeki sokak üzerinde işaretlenmiş ceplere yerleştirileceği varsayılmıştır. Ceplerin hemen yanında ve kaldırım üzerinde şarj için kullanılacak kablolarının yer aldığı bağlantı noktaları yer almaktadır. Buradaki kablolar çekildiğinde içerideki mekanizma vasıtası ile sarımlarından açılarak uzayabilmektedir. Bırakıldığında ise toplanmaktadır. Bu tür sistemler elektrik süpürgelerinin elektrik kablolarında çoğunluklar kullanılmaktadır. Elektrik kabloları dış müdahalelere karşı dayanıklı, kaçak ve kısa devre gibi riskleri ortadan kaldıracak bir yapıda ve dış kaplamalara sahip şekilde tasarlanmıştır. Bu elektrik kabloları her sokağın uç noktalarında yer alan ve karşılıklı yerleştirilmiş terminallere bağlanmaktadır. Bu terminaller ana sunucular ile güvenli bir altyapı ile bağlanmış ve müdahalelere açık değildir. Aradaki iletişim şifrelenmiş ve aktarılan bilginin güvenli olduğu ve değiştirilemediği varsayılmaktadır.

Araç park cebine yanaştığında şarj kablosu çekilerek araca takılacaktır. Takıldıktan sonra kablolu veya kablosuz olmasına bakılmaksızın terminal ile araç arasında doğrulama işlemi sağlanacaktır. Doğrulama işlemi esnasında veya öncesinde araç sahibi terminal üzerinden yüklemek istediği miktarı girerek ve sonrasında ödemesini kredi kartı, nakit vb. yapabilecektir. Ayrıca mobil cihazından veya internete bağlı uyumlu herhangi bir cihaz üzerinden ön tanımlı hesabına bakiye yükleme ve mevcut bakiyesinden aracını şarj edebilme imkânına da sahip olacaktır. İlgili terminaller tüm bu ödeme işlemlerine olanak verecek şekilde tasarlanmıştır. Aynı zamanda terminal şarj bitişinde şarjı kesecek ve sunucu ilgili faturalandırma bilgisini aracın sahibine kısa mesaj, elektronik posta vb. tercihe göre iletecektir. Bunlara ek olarak istenmesi durumunda terminal üzerinde bulunan fiş basma imkânı ile sokakta da fiş basılabilecektir.

Doğrulamanın gerçekleştirilmesi araç ile sunucu arasındaki karşılıklı el sıkışma işlemi olarak değerlendirilebilir. Günlük hayattaki kullanımına paralel olarak el sıkışma olarak tasvir edilmiş doğrulama işlemi ile anlaşma, karşılıklı birbirini tanıma ve güvenme gibi çok önemli işlevler yerine getirilir. Ayrıca iki cihaz arasındaki bu işlemin hangi aşamalardan oluştuğu, hangi taraflara ve işlemlere sahip olduğu gibi bir çerçeve oluşturulması ile bu bir protokole dönüşmektedir. Bu işlem için tasarlanan ve uygulanmasının bilinen saldırılara karşı koruma sağlayacağı varsayılan bir doğrulama protokolü önerilmiştir.

Bu protokolün şeması Şekil 3'te verilmiştir.

*Şekil 3. Doğrulama ve Faturalama Protokolü*

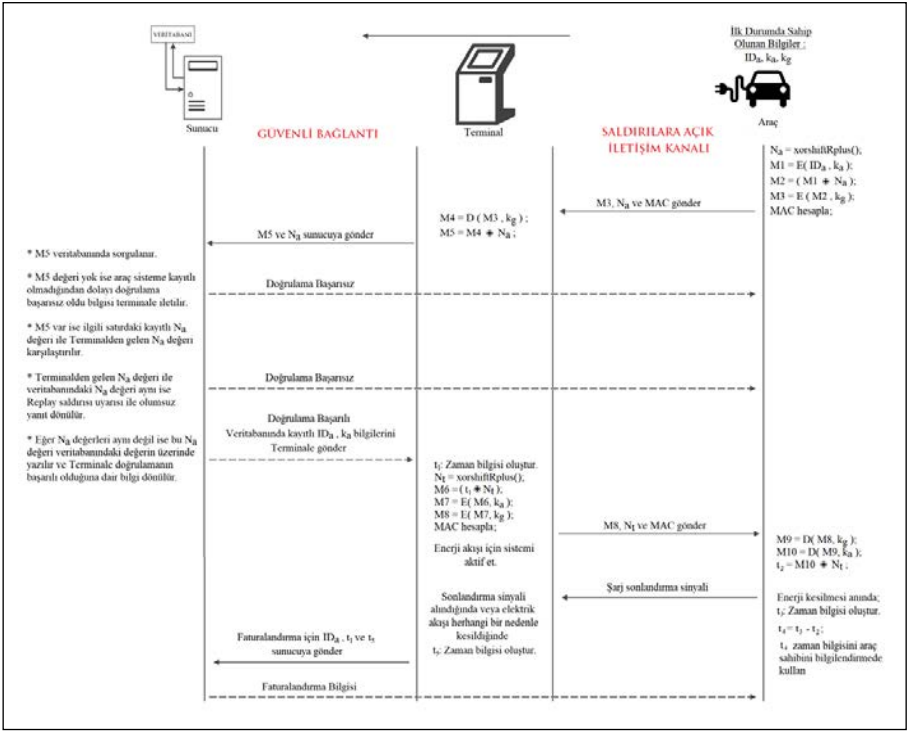

Doğrulama protokolü adımlarını detaylı incelemeden önce sistemin taraflarının sahip olduğu bilgiler ile sistemin durumu hakkındaki genel bilgileri paylaşalım.

Araç üzerinde önceden sunucu üzerinde kayıtları oluşturularak kaydedilmiş IDa, ka ve kg bilgileri bulunmaktadır. Araç ve terminal xorshiftRplus (Çabuk vd., 2017) sözde rastgele sayı üretecini (SRSÜ) kullanarak rastgele sayı üretebilmektedir. Daha düşük kaynaklı sistemler için ise xorshiftULplus (Aydın ve Kösemen, 2020) SRSÜ önerilebilir. Fakat genel olarak araç ve terminal donanım ve bağlantı kaynağı bakımında yeterli kaynaklara sahip olacağından bu çalışmada daha üst güvenlik sağlayan xorshiftRplus kullanılmıştır. Terminal şifreleme de kullanmak üzere kg değerine sahiptir. Bu simge ve gösterimler Tablo 2'de açıklamaları ile verilmiştir.

*Tablo 2. Kullanılan İşaret Ve Simgeler*

| | |
|---|---|
| $k_g$ | Grup üyeleri için gizli grup anahtarı |
| $k_a$ | Araç gizli anahtarı |
| $ID_a$ | Araç tanımlama kodu |
| $N_a$ | Araç tarafında oluşturulan rastgele sayı |
| $N_t$ | Terminal tarafında oluşturulan rastgele sayı |
| MAC | Mesaj doğrulama kodu |
| $\oplus$ | Özel veya işlevi |
| E(X, k) | AES kullanarak X'in k gizli anahtarı ile şifrelenmesi |
| D(E, k) | AES kullanarak X'in k gizli anahtarı ile şifresinin çözümlenmesi |
| t | Zaman değeri |
| M | Hesaplanan mesaj değerleri |

Saldırganın aşağıdaki yeteneklere, erişime, bilgilere ve kaynaklara sahip olduğu varsayılır (Aydın vd., 2020):

- Saldırgan, araç ve terminal arasındaki tüm mesajları dinleyebilir.

- Saldırgan, iletişim kanalında iletilen verileri engelleyebilir.

- Saldırgan karşı tarafa araç veya terminalmiş gibi mesaj gönderebilir.

- Saldırganın gizli parametrelere erişimi yoktur, ancak tüm işlevlere veya işlemlere erişebilir. Sözde rastgele sayı üreteci, şifreleme ve XOR.

- Saldırgan, iletilen tüm mesajları okuyabilir, oluşturabilir, değiştirebilir ve silebilir ve bu mesajları iletişim kanalına geri döndürebilir.

Saldırganın bu yeteneklerinin olduğu varsayımı altında güvenli olduğunu varsaydığımız doğrulama protokolü adımları şu şekildedir.

## Adım 1 (Araç):

- xorshiftRplus SRSÜ kullanarak $N_a$ rastgele sayısı (RS) üretilir.
- Araç tanımlama kodu $ID_a$, $k_a$ anahtarı ile AES algoritması kullanılarak şifrelenir ve M1 oluşturulur.
- M1 ile $N_a$ mantıksal özel veya (xor) işlemine tabi tutulur ve M2 oluşturulur.
- M2, $k_g$ grup gizli şifreleme anahtarı kullanılarak AES ile şifrelenir ve M3 oluşturulur.
- MAC hesaplanır.
- M3, MAC ve $N_a$ terminale gönderilir.

## Adım 2 (Terminal):

- Araçtan iletilen M3, MAC ve $N_a$ alınır.
- M3 değerinin $k_g$ grup gizli anahtarı kullanılarak AES ile şifresi çözülür v M4 elde edilir.
- M4 ile $N_a$ mantıksal özel veya (xor) işlemine tabi tutulur ve M5 oluşturulur.
- M5 ve $N_a$ güvenli bağlantı üzerinden sunucuya iletilir.

## Adım 3 (Sunucu):

- Terminal tarafından güvenli bağlantı üzerinden iletilen M5 ve $N_a$ alınır.
- Veritabanında M5 için arama yapılır.
- Kayıt bulunamaz ise Terminale doğrulamanın başarısız olduğu bilgisi dönülür.
- M5 değeri bulundu ise ilgili veritabanı kaydının $N_a$ değerine bakılır.
- Eğer Terminal üzerinden gelen $N_a$ değeri ile veritabanı üzerindeki $N_a$ değeri aynı ise tekrar(replay) saldırısı olarak değerlendirilir ve doğrulama işlemi başarısız olduğu bilgisi terminal dönülür.

- Eğer terminal üzerinden gelen $N_a$ değeri ile veritabanı üzerindeki $N_a$ değeri aynı değil ise veritabanındaki değer terminalden iletilen yeni değer ile güncellenir ve ilgili veritabanı kaydındaki $ID_a$ ve $k_a$ bilgisi güvenli hat üzerinden terminale gönderilir.

### Adım 4 (Terminal):
- Terminal sunucudan $ID_a$ ve $k_a$ değerlerini alır.
- $t_1$ zaman değerini üzerindeki yerleşik saat yardımı ile alır.
- xorshiftRplus SRSÜ kullanarak $N_t$ rastgele sayısı (RS) üretilir.
- $t_1$ ile $N_t$ mantıksal özel veya (xor) işlemine tabi tutulur ve M6 oluşturulur.
- M6, $k_a$ anahtarı ile AES algoritması kullanılarak şifrelenir ve M7 oluşturulur.
- M7, $k_g$ anahtarı ile AES algoritması kullanılarak şifrelenir ve M8 oluşturulur.
- MAC hesaplanır.
- Enerji akışı aktif edilir.
- M8, MAC ve $N_t$ araca gönderilir.

### Adım 5 (Araç):
- Terminalden gönderilen M8, MAC ve $N_t$ alınır.
- M8 değerinin $k_g$ grup gizli anahtarı kullanılarak AES ile şifresi çözülür ve M9 elde edilir.
- M9 değerinin $k_a$ araç gizli anahtarı kullanılarak AES ile şifresi çözülür ve M10 elde edilir.
- M10 ile $N_t$ mantıksal özel veya (xor) işlemine tabi tutulur ve $t_2$ elde edilir.
- Şarj işleminin sonlanması beklenir.

Enerjinin herhangi bir nedenle kesilmesi durumunda (batarya dolması, dolum bakiyesine ulaşılması, el ile kablo sökülmesi vb.)

### Adım 6 (Araç):

- Araç kendi üzerindeki saat vasıtası ile $t_3$ zaman bilgisini oluşturur.
- $t_3$ zaman değerinden $t_2$ zaman değerini çıkararak $t_4$ zaman değerini hesaplar.
- $t_4$ zaman değerini araç sahibine bilgi amaçlı gösterir ve araçta ilgili yere kaydeder.

### Adım 7 (Terminal):

- $t_5$ zaman bilgisini oluştur ve ilgili araç bilgileri ($ID_a$) ile birlikte başlangıç ($t_1$) ve bitiş zaman($t_5$) bilgisini sunucuya gönder.

### Adım 8 (Sunucu):

- Terminalden gelen araç bilgileri ve zaman bilgileri ile faturalandırma işlemini gerçekleştir ve gerekli bilgilendirmeleri tercih edilen kanallardan araç sahibine ulaştır.

### SONUÇ

Elektik enerjisinin günlük yaşamda kullanımı uzun yıllardır alışılagelmesine rağmen araçlarda kullanımı ancak son yıllarda yaygın hal almaya başlamıştır. Bu yaygınlaşma ile birlikte otomobil, motosiklet, otobüs, kamyonet ve traktör gibi araçların fosil yakıtlarla çalışanlarına ek olarak elektrikle çalışanları da günlük yaşamımızda yaygınlaşmaktadır. Mevcut pil teknolojileri düşünüldüğünde bu cihazların sıklıkla hatta bazılarının her gün şarj edilmesi gerekliliği bulunmaktadır. Özellikle bireysel ve toplu otoparkların yetersiz olduğu bizim gibi ülkelerde şarj işleminin sokakta yapılması bir gereklilik haline gelecektir. Bu şarj işleminin güvenli, kesintisiz ve her yerde yapılabilmesi için sistemler önerilmelidir. Bu konuda herhangi bir standardın bulunmaması da bu konudaki çalışmaların temel motivasyonu olmuştur. Bu çalışmada tüm bu sorunlara çözüm olabilecek kara araçlarının sokakta şarj edilmesine imkân tanıyabilecek bir sistem önerilmiştir. Sistemde cihazın terminal ile birbirini doğrulaması ve faturalamanın güvenli şekilde yapılabilmesi için de bir doğrulama protokolü önerilmiştir. Nihayetinde bir bilgisayar sistemine dönüşen elektrikli araçların doğrulama işlemleri için de bir protokol önerisi gerekli olmuştur.

Gelecekte bu çalışma yerel yönetimler tarafından hayata geçirilip güvenlik testleri yapılabilir ve donanımsal geliştirmeler ile sistem daha iyi hale getirilebilir.

## KAYNAKÇA

## LIST OF PARTICIPATIONS

| Authors | Affiliation | Country | Page |
|---|---|---|---|
| Adem Üzümcü | Ankara Hacı Bayram Veli University | Turkey | 245 |
| Ahmet Nedim Yüksel | Tekirdağ Namık Kemal University | Turkey | 104 |
| Ainur Nogayeva | L.N. Gumilev Eurasian National University | Kazakhstan | 225 |
| Ali Eren Alper | Niğde Ömer Halisdemir University | Turkey | 30 |
| Ali Oğuz Diriöz | TOBB University of Economics and Technology | Turkey | 610 |
| Ali Rıza Dal | Ministry of Transport and Infrastructure | Turkey | 645 |
| Aşkın Özdağoğlu | Dokuz Eylül University | Turkey | 118 |
| Büşra Çiçekalan | Istanbul Technical University | Turkey | 549 |
| Büşra Yılmaz | Aksaray University | Turkey | 491 |
| Cevat Ozarpa | Karabük University | Turkey | 549 |
| Ece Göl | Karamanoğlu Mehmetbey University | Turkey | 440 |
| Eda Nur Erzurum | Konya Technical University | Turkey | 536 |
| Elif Yüksel-Türkboyları | Tekirdağ Namık Kemal University | Turkey | 104 |
| Emine Dilara Aktekin | Niğde Ömer Halisdemir University | Turkey | 129 |
| Emre Esat Topaloglu | Sırnak University | Turkey | 580 |
| Engin Koç | Bursa Technical University | Turkey | 389 |
| Erol Koycu | Sırnak University | Turkey | 580 |
| Fatih Yılmaz | Ministry of Transport and Infrastructure | Turkey | 155, 645 |
| Fatma Nur Doğar | KTO Karatay University | Turkey | 170 |
| Fatma Ünlü | Erciyes University | Turkey | 285, 368 |
| Fikret Müge Alptekin | Ege University | Turkey | 468 |
| Halil İbrahim Kaya | Sivas Cumhuriyet University | Turkey | 403 |
| İsmail Tamboğa | Karamanoğlu Mehmetbey University | Turkey | 440 |
| Kevser Yılmaz | Dokuz Eylül University | Turkey | 118 |

| Authors | Affiliation | Country | Page |
|---|---|---|---|
| Mahmut Suat Delibalta | Niğde Ömer Halisdemir University | Turkey | 2 |
| Mehmet Demiral | Niğde Ömer Halisdemir University | Turkey | 69,129 |
| Melek Çağla Erbil | Istanbul Technical University | Turkey | 549 |
| Melih Soner Çeliktaş | Ege University | Turkey | 468 |
| Melisa Arslan | Muğla Sıtkı Koçman University | Turkey | 202 |
| Mustafa Uslu | Düzce University | Turkey | 40 |
| Mustafa Yasir Kurt | Social Sciences University of Ankara | Turkey | 336 |
| Oğuz Kara | Düzce University | Turkey | 40 |
| Ömer Aydın | Dokuz Eylül University | Turkey | 305 |
| Özge Demiral | Niğde Ömer Halisdemir University | Turkey | 69 |
| Özlem Fındık Alper | Niğde Ömer Halisdemir University | Turkey | 30 |
| Pelin Gençoğlu | Erciyes University | Turkey | 285,368 |
| Sefa Coşkun | MEF University | Turkey | 317 |
| Selcen Kaçar | KTO Karatay University | Turkey | 170 |
| Selçuk Sayın | Konya Technical University | Turkey | 536 |
| Selim Şanlısoy | Dokuz Eylül University | Turkey | 632 |
| Sevda Kuşkaya | Erciyes University | Turkey | 368 |
| Sinan Erdoğan | Hatay Mustafa Kemal University | Turkey | 19 |
| Sinem Atıcı Ustalar | Atatürk University | Turkey | 632 |
| Soner Yakar | Çukurova University | Turkey | 491 |
| Şerife Özkan Nesimioğlu | KTO Karatay University | Turkey | 170 |
| Tuğçenur Ekinci Furtana | İstanbul Ticaret University | Turkey | 563 |
| Yunus Beyhan | MEF University | Turkey | 317 |
| Zeynep Paralı | Adnan Menderes University | Turkey | 517 |
| Zoran Ivanov | TOBB University of Economics and Technology | Turkey | 610 |
| Жайлыбаев Дәулет | Eurasian Research Institute | Kazakhstan | 197 |